\documentclass[a4paper,twocolumn,10pt,accepted=2025-04-10]{quantumarticle}
\pdfoutput=1
\usepackage[mathletters]{ucs} 
\usepackage[utf8x]{inputenc}
\usepackage[english]{babel}
\usepackage{hyperref}
\usepackage{tikz}
\usepackage{lipsum}
\usepackage{cite}
\usepackage{graphicx}
\usepackage{amsmath,amssymb}
\usepackage{dcolumn}
\usepackage{mathrsfs}
\usepackage{physics}
\usepackage{float}
\makeatletter
\let\newfloat\newfloat@ltx
\makeatother
\usepackage{booktabs}
\usepackage{svg}
\usepackage{comment}
\usepackage[frozencache,cachedir=.]{minted}

\usepackage[textsize=tiny,textwidth=2.2cm,bordercolor=white]{todonotes}

\definecolor{customcolorblue}{HTML}{4573ae}

\definecolor{LightGray}{gray}{0.9}

\newcommand{\code}[1]{{\fontfamily{pcr}\selectfont #1}}

\begin{document}

\title{WaveguideQED.jl: An Efficient Framework for Simulating Non-Markovian Waveguide Quantum Electrodynamics}

\author{Matias Bundgaard-Nielsen}
\affiliation{Department of Electrical and Photonics Engineering, Technical University of Denmark, Building 343, Lyngby, Denmark}
\orcid{0000-0003-2941-6643}
\affiliation{NanoPhoton-Center for Nanophotonics, Technical University of Denmark, Building 343, Lyngby, Denmark}
\author{Dirk Englund}
\orcid{0000-0002-1043-3489}
\affiliation{Department of Electrical Engineering and Computer Science, Massachusetts Institute of Technology, Cambridge, MA 02139, USA}
\author{Mikkel Heuck}
\orcid{0000-0001-9769-6005}
%\email{mheu@dtu.dk}
\affiliation{Department of Electrical and Photonics Engineering, Technical University of Denmark, Building 343, Lyngby, Denmark}
\affiliation{NanoPhoton-Center for Nanophotonics, Technical University of Denmark, Building 343, Lyngby, Denmark}
\author{Stefan Krastanov}
\orcid{0000-0001-5550-5258}
\affiliation{Manning College of Information and Computer Sciences, University of Massachusetts Amherst, 140 Governors Drive
Amherst, Massachusetts 01003, USA}
\email{skrastanov@umass.edu}

\maketitle

\begin{abstract}
In this paper, we introduce a numerical framework designed to solve problems within the emerging field of Waveguide Quantum Electrodynamics (WQED). The framework is based on collision quantum optics, where a localized quantum system interacts sequentially with individual time-bin modes. This approach provides a physically intuitive model that allows researchers familiar with tools such as QuTiP in Python, Quantum Optics Toolbox for Matlab, or QuantumOptics.jl in Julia to efficiently set up and execute WQED simulations. Despite its conceptual simplicity, we demonstrate the framework’s robust ability to handle complex WQED scenarios. These applications include the scattering of single- or two-photon pulses by quantum emitters or cavities, as well as the exploration of non-Markovian dynamics, where emitted photons are reflected back, thereby introducing feedback mechanisms. 
\end{abstract}

\hyperlink{https://github.com/qojulia/WaveguideQED.jl}{https://github.com/qojulia/WaveguideQED.jl}  
%\todo{Can we shoehorn "non markovian" in the title? I am wondering whether we should remove the package name? ANSWER: The quantum cumulants paper has the name in the title.}
% -------------------------------------------------------
\section{Introduction \label{sec:intro}}
% -------------------------------------------------------
Photons transporting quantum information or mediating entanglement between localized quantum systems is one of the promising backbones of quantum technology. Here, the temporal shape of the flying photons is crucial and warrants a multimode description \cite{Shapiro2006Single-photonComputation,Heuck2020Controlled-PhaseNonlinearities,Sheremet2023WaveguideCorrelations}.

%A key ingredient is here the ability to have photon-photon interactions.  However, the combination of low optical power and weak nonlinearities in bulk materials has proven a significant challenge in achieving high-fidelity interactions between quantum states of light\cite{Chang2014QuantumPhoton}. 

Recent progress in the design and fabrication of nanostructures holds promise for an efficient interface with photons. Optical waveguides allow for the guiding and manipulation of photons with the possibility for on-chip scalability \cite{Lodahl2015InterfacingNanostructures}. 
By integrating a localized emitter in the waveguide, the light-matter interaction can mediate photon-photon interactions, leading to optical nonlinearities \cite{Chang2014QuantumPhoton}. Indeed, an efficient light-matter interface in a photonic crystal waveguide with a semiconductor quantum dot has been demonstrated \cite{LeJeannic2022DynamicalEmitter}. Traveling photon states are also found in superconducting qubits \cite{Grebel2024BidirectionalNodes}, and even traveling photon-phonon translators have been proposed \cite{Safavi-Naeini2011ProposalTranslator}. While the Jaynes-Cummings model effectively describes the interaction between a localized optical mode and a quantum emitter \cite{Jaynes1963ComparisonMaser}, the multimode nature of traveling photonic states interacting with a localized quantum system necessitates extensions, requiring more sophisticated theoretical frameworks.

%While the Jaynes-Cummings model aptly describes the interaction between a localized optical mode and a quantum emitter \cite{Jaynes1963ComparisonMaser}, the interaction between a traveling photonic state and a localized quantum system is multimode in nature and requires more careful treatment.
\begin{table*}[htbp]
\centering
\caption{Comparison of selected approaches for solving WQED problems}
\label{tab:methods_comparison}
\small
\begin{tabular}{|l|c|c|c|c|c|}
\hline
\textbf{Approach} & \textbf{\begin{tabular}[c]{@{}c@{}} \# of \\ photons \end{tabular}} & \textbf{\begin{tabular}[c]{@{}c@{}}Includes \\ Feedback\end{tabular}} & \textbf{\begin{tabular}[c]{@{}c@{}}Output\end{tabular}} & \textbf{\begin{tabular}[c]{@{}c@{}}Input Type\end{tabular}} & \textbf{\begin{tabular}[c]{@{}c@{}}Availability\end{tabular}} \\ \hline
Input-Output Theory (SLH)\cite{Kiilerich2019Input-OutputPulses,Kiilerich2020QuantumRadiation}      & Several  & No & Observables & Product states & Custom \\ \hline
Tensor Networks\cite{Pichler2016PhotonicFeedback,Guimond2017DelayedPerspective,ArranzRegidor2021ModelingModel}                   & Several  & Yes & Full state & Arbitrary state & Custom \\ \hline
WaveguideQED (This Work) & Up to 2  & Yes & Full state & Arbitrary state & Open Source \\ \hline
Stochastic Master Equations\cite{Baragiola2012N-PhotonSystem,Baragiola2017QuantumStates}   & Several & No & Observables & Product states & Custom \\ \hline
Scattering Matrix Formalisms\cite{Fischer2018ScatteringSystem, Pletyukhov2015QuantumEntropy}   & Up to 2  & No & Observables & Coherent pumping & Open Source \\ \hline 
Hierarchical equations of motion \cite{heomfuchs2023,liang2024purifiedinputoutputpseudomodemodel,cirio2025inputoutputhierarchicalequationsmotion} & Several & Yes & Observables & Arbitrary state & Custom \\ \hline
\end{tabular}
\end{table*}

Although numerous simulation tools exist for modeling the dynamics of open cavity quantum electrodynamical systems, where the localized optical mode is often described by a master equation, there are limited options available for simulating and including quantum mechanical states of traveling photons (see table \ref{tab:methods_comparison}). Given that one of the most promising applications of photons is their ability to transmit or mediate quantum information over long distances, a precise and accurate description of their traveling states is crucial. Thus, the field of Waveguide Quantum Electrodynamics (WQED) is an exciting new area of physics that has recently received much attention, including protocols for producing 2D-cluster states \cite{Pichler2016PhotonicFeedback,Pichler2017UniversalFeedback}, efficient light-matter interaction with quantum emitters embedded in photonic crystal waveguides \cite{LeJeannic2021ExperimentalWaveguide,LeJeannic2022DynamicalEmitter}, scattering of quantum pulses on Fano resonances \cite{Xu2016FanoTransport,Joanesarson2020Few-photonGeometries}, as well as quantum feedback \cite{Whalen2017OpenFeedback,Barkemeyer2021StronglyStateb}. The theory of traveling quantum states of light is inherently a rich and complex many-body problem, where solutions can be obtained using e.g.\ path-integrals \cite{Shi2015Multiphoton-scatteringEquations}, input-output theory and scattering matrices \cite{Caneva2015QuantumFormalism,Pletyukhov2015QuantumEntropy,Trivedi2018Few-photonSystems,Fischer2018ParticleSystems}, Feynman diagrams \cite{Roulet2016SolvingComputation}, equations of motion for the wavefunction \cite{Nysteen2015ScatteringCorrelations,Konyk2016QuantumWaveguide}, or matrix-product states \cite{Guimond2017DelayedPerspective}.

A non-exhaustive list applying some of the above methods is shown in table \ref{tab:methods_comparison}. Here, stochastic master equations \cite{Baragiola2012N-PhotonSystem,Baragiola2017QuantumStates}, input-output theory (SLH) \cite{Kiilerich2019Input-OutputPulses,Kiilerich2020QuantumRadiation,Lund2023PerfectPulse,Yang2022DeterministicSystems}, or scattering matrix-formalisms \cite{Fischer2018ScatteringSystem} all correctly predict the interaction of wavepackets with cavity-emitter systems, but in these descriptions, the complete state of the traveling wavepacket is not described. Treating non-Markovian effects, such as delayed feedback, is therefore extremely difficult, and only photon states without time entanglement can be included.  Recently, Hierarchical equations of motion (HEOM) \cite{heomfuchs2023,liang2024purifiedinputoutputpseudomodemodel,cirio2025inputoutputhierarchicalequationsmotion}, have been extended to include non-Markovian effects in waveguides while also allowing non-flat spectral densities. Only observables of the environment are, however, available, and also the generality comes at the cost of increased numerical complexity. To treat such problems, matrix product states can be used to describe the complete state of the waveguide in an efficient way \cite{Pichler2016PhotonicFeedback,Pichler2017UniversalFeedback,ArranzRegidor2021ModelingModel,Richter2022EnhancedNetworks}. Matrix product states are, however, complex and offer little physical insight into the system. Alternative models using a more intuitive space-discretization of the waveguides have, therefore, been proposed \cite{Crowder2020QuantumFeedback,Crowder2022QuantumFeedback}. Many of the aforementioned approaches, however, only exist as ad-hoc in-house numerical implementations, rarely of high performance, and engineered as non-generalizable solutions to one-off problems. 

Here, we present WaveguideQED.jl, the first open-source high-performance realization of the quantum collision model \cite{Ciccarello2018CollisionOptics}. It describes the traveling wavepacket as a collection of time-bins interacting with quantum systems one at a time \cite{Heuck2020Photon-photonCavities,Heuck2020Controlled-PhaseNonlinearities,Krastanov2022Controlled-phaseEmitter}. We introduce techniques for the auto-generation of appropriate basis for the traveling multi-photon wavepackets, and utilize a number of already known but not widely applied techniques in high-performance non-allocating application of highly-structured time-dependent operators. The framework is conceptually simple and easy to use for researchers familiar with quantum optics simulation software such as QuTiP in python \cite{Johansson2012QuTiP:Systems,Johansson2013QuTiPSystems}, Quantum Optics Toolbox in Matlab, and QuantumOptics.jl in Julia \cite{Kramer2018QuantumOptics.jl:Systems} as wel as QuantumToolbox.jl in Julia. In fact, it is compatible with  QuantumOptics.jl, and arbitrary local quantum systems defined in QuantumOptics.jl can be combined with waveguide states and operators defined in WaveguideQED.jl. The framework serves as an entry point to the field of WQED for less experienced researchers and as a simulation tool for more experienced researchers who want to explore non-Markovian feedback effects or complex two-photon entangled states.

We demonstrate the capabilities of the framework by investigating the scattering of single and two-photon pulses off emitters coupled to single and multiple waveguide channels. Despite the simplicity of these two examples, the theory of computing the scattering matrices is quite involved, and we demonstrate how easy it is to simulate and describe using our framework. To further highlight the power of the framework, we consider non-Markovian effects that arise due to delayed feedback. As an example, we consider a semi-infinite waveguide with a mirror at one end leading to a feedback loop, and, for certain conditions, excitation trapping occurs.  

% -------------------------------------------------------
\section{Theoretical Background \label{sec:theory}}
% -------------------------------------------------------

In the following, we briefly outline the theory and assumptions behind collision quantum optics; for a thorough review, see ref.~\cite{Ciccarello2022QuantumCollisionModels}. We then introduce the discretized time-bin formalism, which the WaveguideQED.jl implementation is based on. We also provide an example of how to calculate the scattering of a single photon pulse by hand using the time-bin formalism, which motivates the introduction of the numerical framework.

In its most general form, we can write the state of a traveling one-photon pulse as \cite{Ciccarello2018CollisionOptics}: 
\begin{equation}
    |\psi\rangle=\int \mathrm{d} \nu \psi(\nu) w^{\dagger}(\nu)| \emptyset \rangle,
\end{equation}
where $w^{\dagger}(\nu)$ is the creation operator for a photon of frequency $\nu$ with units of $\sqrt{\mathrm{time}}$ and $\psi(\nu)$ defines the wavefunction of the pulse also with units of $\sqrt{\mathrm{time}}$. $\ket{\emptyset}$ denotes the vacuum state, i.e., no photon in the waveguide. The free evolution of the pulse is then given by the Hamiltonian:
\begin{equation}
    H_\text{f} = \hbar \int \mathrm{d} \nu \nu w^{\dagger}(\nu) w(\nu)
\end{equation}
The interaction with a localized system, such as a two-level system with the atomic transition operator $\sigma = \ket{g}\bra{e}$ (one could equivalently have considered a localized optical mode with annihilation operator $a$), is then given by \cite{Ciccarello2018CollisionOptics}:
\begin{equation}
    H_\text{int} = \hbar \int \mathrm{d} \nu  \left(g(\nu) \sigma^{\dagger} w(\nu) + g(\nu)^{*} \sigma w^{\dagger}(\nu)\right)
\end{equation}
here $g(\nu)$ defines the coupling strength between the emitter and each individual mode. If we assume $g(\nu) = i \sqrt{\gamma/2\pi}$, a considerable simplification can be made to the above by moving into the interaction picture:
\begin{equation}
    H_\text{time}(t) =  i \hbar \sqrt{\gamma} (\sigma^\dagger w(t)-\sigma w^\dagger(t) )
\end{equation}
where we defined $H_\text{time}(t) = \mathrm{e}^{i H_\text{f} t /\hbar} H_\text{int} \mathrm{e}^{-i H_\text{f} t /\hbar}$ and introduced $w(t)=\int \mathrm{d} \nu w(\nu) e^{-i \nu t}/\sqrt{2 \pi}$ with units of $1/\sqrt{\mathrm{time}}$. The single photon state can equivalently be defined as:
\begin{equation}
    \ket{\psi} = \int_{t_0}^{t_{N}} \mathrm{d}t \ \xi^{(1)}(t) w^\dagger(t) \ket{\emptyset} \label{eq:singlephoton}
\end{equation}
here $\xi^{(1)}(t) = \frac{1}{\sqrt{2 \pi}} \int \mathrm{d} \nu \psi(\nu) e^{-i \nu t} $ is the wavefunction in the time domain. The inner product $\bra{\psi} w^\dagger(t) w(t) \ket{\psi} = |\xi^{(1)}(t)|^2$ with units of inverse time thus gives the flux of photons at time $t$, and multiplied with a small timestep $\Delta t |\xi^{(1)}(t)|^2$ the probability of observing a photon in the interval $[t, t+\Delta t]$.

We can similarly define a two-photon state as:
\begin{equation}
    \ket{\psi} =  \frac{1}{\sqrt{2}} \int_{t_0}^{t_{N}} \!\!d t^{\prime} \int_{t_0}^{t_{N}} \!\!d t \ \xi^{(2)}(t,t^{\prime}) w^\dagger(t) w^\dagger\left(t^{\prime}\right)|\emptyset \rangle  
\end{equation}
where the two-photon wavefunction $\Delta t \abs{\xi^{(2)}(t,t^\prime)}^2$ now represents the probability of observing a photon between $t$ and $t+\Delta t$ and another photon between $t^\prime$ and $t^\prime+\Delta t$. 

\begin{figure}[!ht]
    \centering
    \includegraphics[width = \linewidth]{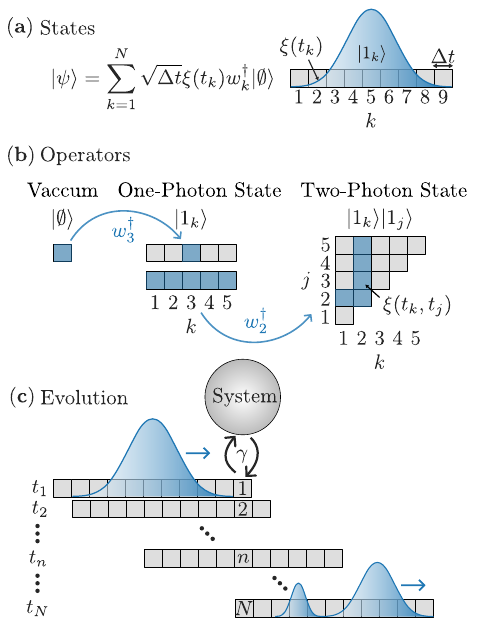}
    \caption{(a): Numerical representation of a time-binned single photon state in WaveguideQED.jl. (b): A discretized waveguide operator $w_3^\dagger$ acts on a vacuum state and creates a single photon in time-bin $k=3$. $w_2^\dagger$ then acts on a single photon state, creating a two-photon state in all bins with an index $j=2$ or $k=2$. (c): A single photon state impinges on a quantum system, and its evolution is calculated by letting each time-bin interact with the system one at a time.}
    \label{fig:conveyor}
\end{figure}

In this picture, the interaction with the localized system is only governed by a single (time) mode $w(t)$ at each time $t$. The complexity of the problem has now been moved from interacting with a collection of bosonic modes to a time-dependent Hamiltonian. As we will see, this allows us to define time-discretized bins that interact with the localized system one at a time. We note that if linear dispersion is assumed, one can directly map from time-binned modes to spatial modes with $t\rightarrow x/c$, where x is the position of the photon and $c$ is the speed of light. Thus, one can simply think of the evolution of the time-binned modes as a moving conveyor belt (as also shown in Fig.~\ref{fig:conveyor}c). By exploiting Lazy operators (a feature of QuantumOptics.jl) and multiple dispatch in Julia, the time-dependent waveguide operators representing the interaction can be implemented as non-allocated kernel functions, as opposed to sparse matrices. In addition, we make an appropriate choice of basis for the waveguide state. All of these performance-enhancing features allow us to have a complete description of the waveguide quantum state, while maintaining an efficient framework. More details on performance are provided in Section~\ref{sec:performance}.

\subsection{Discretization}
In this section, we will introduce the discretized time-bin formalism and use it to derive the scattered single-photon pulse from an emitter. We start by discretizing the continuous Fock state into time-bins of width $\Delta t$. This amounts to defining new discretized annihilation and creation operators as \cite{Heuck2020Photon-photonCavities}:
\begin{equation}
     w(t_k) = w(k \Delta t) \rightarrow  \frac{w_k}{\sqrt{\Delta t}} \ \ \  \text{with} \ \comm{w_j}{w_k^\dagger} = \delta_{jk} 
\end{equation}
where $w_k$ is the descritized (unitless) operator of the $k$th time-bin and the factor of $1/\sqrt{\Delta t}$ assures the commutator relation in the limit of $\Delta t \rightarrow 0$. This means that the single photon continuous Fock state becomes:
\begin{equation}
    \ket{\psi} = \sum_{k=1}^N \sqrt{\Delta t} \xi(t_k) w_k^\dagger \ket{\emptyset} = \sum_{k=1}^N \sqrt{\Delta t} \xi(t_k) \ket{1_k} \label{eq:single_discretized}
\end{equation}
where we defined the time-bin state $w_k^\dagger \ket{\emptyset} = \ket{1_k}$ and the sum runs til $N$, such that $N \Delta t$ is the simulation time. The two-photon state similarly becomes:
\begin{equation}
\begin{aligned}
     \ket{\psi}  &= \sum_{i=1}^N \Delta t \xi^{(2)}\left (t_i,t_i\right )\left|2_i\right\rangle \\
    &+ \frac{1}{\sqrt{2}} \sum_{i=1}^N \sum_{k > i}^N \Delta t (\xi^{(2)}(t_i,t_k) + \xi^{(2)}(t_k,t_i )) \mid 1_{i} 1_{k}\rangle \label{eq:two_discretized}
\end{aligned}
\end{equation}
where we used $\ket{1_{i}1_{k}} = \ket{1_{k}1_{i}}$ and $w_i^\dagger w_i^\dagger \ket{\emptyset} = \sqrt{2} \ket{2_i}$. In Fig.~\ref{fig:conveyor}(a) and (b), we illustrate the numerical representation of the single photon state and the state resulting from the action of the waveguide operator of time-bin $k=3$, $w_3^\dagger$, on a vacuum state. We also show the resulting two-photon state from applying $w_2^\dagger$ to a single photon state.

The added complexity and scaling of more photons is clear, and in general, we will require $\approx N^n$ elements, where $n$ is the number of photons. In WaveguideQED.jl, the maximum number of photons is limited to $n$=2. 
% we limit our considerations to a maximum of $n=2$ photons, which we will see is still enough to investigate the rich dynamics of WQED. 
We discuss the limitations and numerical considerations of this further in Section~\ref{sec:performance}. 
%The cost of describing the whole waveguide state is thus clear, and it is obvious that we are paying a steep price for having access to the whole waveguide state. Luckily, many of the fundamental photon-photon interactions, such as the beamsplitter, can be studied with only a two-photon description. Furthermore, this full description allows us to describe two-photon states of arbitrary complexity with no restriction on the amount of entanglement 
%Although it might be possible to extend the framework to a higher number of photon states and still have acceptable performance, the scaling quickly makes  
%this might not be numerically viable and  
    
The discrete version of the interaction Hamiltonian is:
\begin{equation}
    H_\text{time}(t) =  \sum_k f_k(t) i \hbar \sqrt{\gamma/\Delta t} (\sigma^\dagger w_k - \sigma w_k^\dagger) \label{eq:discretized_interaction}
\end{equation}
where we defined:
\begin{equation}
f_k(t) = \begin{cases}
          1, & \text{if } t_k < t <t_k+\Delta t  \\
          0, & \text{otherwise}
\end{cases}  \label{eq:fk}    
\end{equation}
The Hamiltonian is thus constant within each time-bin and allows for an intuitive mental picture: We are moving a conveyor belt of photon bins and letting one bin of the belt interact with the system at a time. This is also illustrated in Fig.~\ref{fig:conveyor}(c).

\subsection{Scattering off an Emitter}
In the previous section, we introduced the time-bin formalism, and in this section, we are going to use it to derive the dynamics of a single-photon pulse scattering off an emitter. This serves two purposes. First, it gives an idea of the inner workings of WaveguideQED.jl, and second, it motivates the numerical implementation.

The differential equations governing the dynamics are defined from the Hamiltonian in Eq.~\eqref{eq:discretized_interaction} and Schrödinger's equation $i\hbar \partial_t \ket{\psi} = H \ket{\psi}$. Since the Hamiltonian is constant within each time-bin, we can describe the evolution from time-bin $n-1$ to $n$ by the unitary evolution \cite{Heuck2020Photon-photonCavities}:
\begin{equation}
    U_n = U(t_{n-1},t_{n}) = \exp \left(- \int_{t_{n-1}}^{t_{n-1}+\Delta t} \frac{i}{\hbar} H_\text{int}(t^{\prime}) d t^{\prime} \right)
\end{equation}
which relates the state at time $t_{n-1}$ to the state at time $t_n$ as: $U_n \ket{\psi_{n-1}} = \ket{\psi_n}$, where $\ket{\psi_k}$ denotes the state at time $t_k$. To first order in $\Delta t$, $U_n$ is given as:
\begin{equation}
    U_n \approx 1+\sqrt{\gamma \Delta t}\left(a^{\dagger} w_n-a w_n^{\dagger}\right)-\frac{\gamma}{2} \Delta t a^{\dagger} a w_n w_n^{\dagger} \label{eq:unitaryevolution}
\end{equation}

We consider the initial state, where the emitter is in the ground state $\ket{g}$, and the waveguide contains a single-photon wavepacket with wavefunction $\xi(t)$:
\begin{equation}
    \left|\psi_0\right\rangle = \sum_{k=1}^N \sqrt{\Delta t} \xi(t_k) \ket{g} \ket{1_k} \label{eq:initial_single}
\end{equation}

By applying $U_n$ recursively on Eq.~\eqref{eq:initial_single}, we find the state $\ket{\psi_{n}}$ to be:
\begin{equation}
\begin{aligned}
\ket{\psi_n} = & \sum_{k=n+1}^N \xi(t_k) \sqrt{\Delta t} \ket{g}\ket{1_k}+ \\
& \sum_{k=1}^n \xi_\text{out}(t_k) \sqrt{\Delta t}|g\rangle\left|1_k\right\rangle+\psi_e(n)|e\rangle|\emptyset\rangle
\end{aligned}
\end{equation}
After $n$ time steps, the state has three distinct terms. The first term is the waveguide field, which has not yet interacted with the emitter. The second term is the waveguide field after it has interacted with the emitter, and we label this $\xi_\text{out}$, which eventually will be the scattered field. The third term, $\psi_e(n)$, describes the excitation probability of the emitter at time step $n$. Relating $\psi_e(n)$ to $\psi_e(n-1)$, we find the following equation of motion \cite{Heuck2020Photon-photonCavities}:    
\begin{equation}
    \frac{d \psi_e(t)}{d t}  = -\frac{\gamma}{2} \psi_e(t) + \sqrt{\gamma} \xi(t) \label{eq:eom1}
\end{equation}
and the well-known input-output relation:
\begin{equation}
    \xi_\text{out}(t_{n+1}) = \xi(t_{n+1}) - \sqrt{\gamma} \psi_e(t_{n}). \label{eq:eom2}
\end{equation}
In this case, the derivation of the scattered field was quite simple, but one could easily imagine more complicated scenarios. Such scenarios could include describing additional quantum systems coupled to the local emitter, additional terms in the Hamiltonian, or simply just a detuning between the carrier frequency of the waveguide pulse and the emitter. Even a different initial condition, containing, e.g., a two-photon pulse, would greatly complicate things. In all of these examples, one would have to go through the derivation again, with increasing complexity. See, for example, Ref.~\cite{Heuck2020Photon-photonCavities}, where an initial state containing two photons in the waveguide is impinging on a cavity containing two modes and a  $\chi^{(3)}$ nonlinearity. Here, multiple photon branches need to be considered in the recursive derivation, which becomes quite lengthy.

Instead, the idea of WaveguideQED.jl is to represent the waveguide state and the associated operators as efficiently as possible to allow for a numerical solution of the differential equation formed by the Schrödinger equation. As mentioned, this is achieved by using a custom sparse basis for the waveguide state and implementing operators on this basis as compact function kernels instead of sparse matrices. The operators and states can still be effortlessly combined with arbitrary operators and states defined already in QuantumOptics.jl. Interactions between the waveguide and complex local systems can thus be studied easily. This is a major advantage of the framework and is a clear distinction between other approaches, where there might be complexities involved in deriving the dynamics of the local system. In the following section, we will introduce the components of the framework. 

% -------------------------------------------------------
\section{WaveguideQED.jl Framework \label{sec:framework}}
% -------------------------------------------------------
In the last section, we introduced the time-bin formalism for calculating the dynamics of interactions between photons in waveguides and local stationary quantum systems. Motivated by the tedious derivation of the equations of motion for the combined system of the waveguide and localized system, we instead seek to represent the waveguide operators $w_k$ numerically to construct the Hamiltonian in Eq.~\eqref{eq:discretized_interaction} and likewise represent states such as the ones defined in Eq.~\eqref{eq:single_discretized} and \eqref{eq:two_discretized}. In this section, we introduce the framework's components and provide a basic example of how to use them. The waveguideQED.jl package works as an extension to the QuantumOptics.jl package \cite{Kramer2018QuantumOptics.jl:Systems} and relies on much of the infrastructure introduced and provided in QuantumOptics.jl. This means that users familiar with QuantumOptics.jl will find it simple and intuitive to include the three building blocks of WagveguideQED.jl: the waveguide basis, the waveguide operators, and the waveguide states in their simulations. For more details and instructions on using the WaveguideQED.jl package, see the documentation \cite{DocumentationWaveguideQED.jl} or the publicly available source code hosted on Github~\cite{GithubWaveguideQED.jl}. 

\subsection{Waveguide Basis and Operators \label{subsec:waveguidebasis}}

The waveguide basis is the object from which the user defines waveguide states (kets) and operators. The basis contains information about the number of bins the waveguide contains, which is used when waveguide states and operators are combined with operators of different Hilbert spaces. This allows the user to build the total quantum system through tensor products. In Code Sample \ref{ls:bando}, we define a waveguide basis containing a single-photon waveguide state that spans over the time range $t=0$ to $t=10$ with time steps of $\Delta t = 0.05$ (denoted \code{dt} in the code). From the basis \code{bw}, we define the waveguide operators and combine them with the raising and lowering operators of a two-level system to obtain the Hamiltonian in Eq.~\eqref{eq:discretized_interaction}. Note that a Fock basis with a cutoff of one photon is equivalent to a two-level system.

\begin{listing}[!ht]
\begin{minted}[
framesep=2mm,
baselinestretch=0.9,
bgcolor=LightGray,
fontsize=\small,
]{julia}
#Load packages
using WaveguideQED
using QuantumOptics

#Define the simulation time and basis
times = 0:0.05:10
dt = times[2] - times[1]
bw = WaveguideBasis(1,times)

#Basis of two-level-system
be = FockBasis(1)

#Define waveguide operators
w = destroy(bw)
wd = create(bw)

#Rasing and lowering operators
s = destroy(be)
sd = create(be)

#Hamiltonian
γ = 1.0
H = im*sqrt(γ/dt)*(sd ⊗ w - s ⊗ wd)
\end{minted}
\caption{Waveguide bases and operators are defined and combined with operators belonging to other Hilbert spaces. Here, $\otimes$ is the tensor product and can written in Julia with: \texttt{\textbackslash otimes} \label{ls:bando}}
\end{listing}

The time dependence of the waveguide operators \code{w} and \code{wd} are here hidden and taken care of behind the scenes when the Hamiltonian is passed to the solver. The waveguide operators are not implemented as matrices but instead as kernel functions that perform the relevant waveguide operator effect. This enables a very efficient implementation of the time dependence that leverages Julia's multiple dispatch and the Lazy operators implemented in QuantumOptics.jl. We discuss the performance advantage of this ``matrix-free'' multiplication further in Section~\ref{sec:performance}. The cost of having a matrix-free multiplication is, however, that the tensor product between operators cannot be performed trivially, and instead, we use Lazy Operators \cite{LazyOperatorsQuantumOptics.jl}. Lazy Operators delay the evaluation of the tensor product until the actual multiplication, which, in this case, is a much more efficient implementation of the time dependence. It is the almost semi-symbolic nature of Lazy Operators that enables us to use matrix-free kernel functions to perform the non-allocating application of the time-dependent waveguide operators, thus greatly increasing efficiency. %\todo{it is more efficient because... maybe also explain a bit more how Lazy operators are basically semi-symbolic} 
%The waveguide basis can also be defined to include two-photon states and multiple waveguide modes (still with the limitation of a total of two photons being present). See section ?? and ?? for examples where multiple photons are described. 

\subsection{Waveguide States and Simulation}
Having defined the Hamiltonian in the previous section, we are now ready to simulate the scattering of a single-photon pulse off an emitter. We need to define the initial waveguide state in Eq.~\eqref{eq:single_discretized}. This is done in Code Sample \ref{ls:single_photon_scattering}, where the single photon has a Gaussian shape. The waveguide state is combined with the emitter state, initially in the ground state, and the scattering is calculated by solving the Schrodinger equation defined using the Hamiltonian in Code Sample~\ref{ls:bando}.

\begin{listing}[!ht]
\begin{minted}[
framesep=2mm,
baselinestretch=0.9,
bgcolor=LightGray,
fontsize=\small,
]{julia}
#Define gaussian input state
ξ(t,τG,t0) = sqrt(2/τG)*(log(2)/pi)^(1/4)*
exp(-2*log(2)*(t-t0)^2/τG^2)
τG,t0 = 1,5

#Define input state
ψ_w = onephoton(bw,ξ,τG,t0)
ψ_in = fockstate(be,0) ⊗ ψ_w

#Calculate output state
sol = waveguide_evolution(times,ψ_in,H)
ψ_out = OnePhotonView(sol)
\end{minted}
\caption{An initial single-photon state with a Gaussian shape is defined. The waveguide state is combined with an emitter initially in the ground state. The scattered state \code{ψ\_out} is calculated from the initial state \code{ψ\_in} and the Hamiltonian from Code Sample \ref{ls:bando}. The scattered state \code{ψ\_out} is viewed using
\code{OnePhotonView} and plotted in Fig.~\ref{fig:single_photon_scattering}. }
\label{ls:single_photon_scattering}
\end{listing}

\begin{figure}
    \centering
    \includegraphics[width=\linewidth]{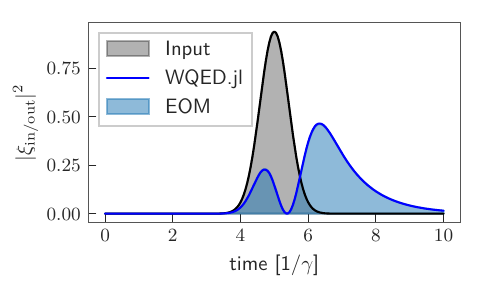}
    \caption{The wavefunction of the single photon state as a function of time before and after scattering off the emitter. The wavefunction is calculated using the equation of motion (EOM) in eqs.~\eqref{eq:eom1} and \eqref{eq:eom2} and WaveguideQED.jl, and the two approaches are seen to agree. Parameters are the same as in Code Sample \ref{ls:single_photon_scattering}.}
    \label{fig:single_photon_scattering}
\end{figure}

In Fig.~\ref{fig:single_photon_scattering}, we plot the scattered wavefunction obtained from Code Sample \ref{ls:single_photon_scattering} together with the solution to the equations of motion (EOMs) in Eqs.~\eqref{eq:eom1} and \eqref{eq:eom2}. We see that the two solutions agreement agree very well. In Appendix \ref{app:single_photon}, we show convergence of the deviation between the EOM and WaveguideQED.jl solutions as the number of time-bins used to represent the waveguide state is increased (or equivalently $\Delta t$ is decreased). Generally, the time-bins should be small enough to capture the fastest timescale of the system so that the wavefunction is quasi-constant during each bin.

\begin{listing}[!ht]
\begin{minted}[
framesep=2mm,
baselinestretch=0.9,
bgcolor=LightGray,
fontsize=\small,
]{julia}
#Waveguide basis with two photons
bw = WaveguideBasis(2,times)
w,wd = destroy(bw),create(bw)
H = im*sqrt(γ/dt)*(sd ⊗ w - s ⊗ wd)

#Define twophoton gaussian input state
ξ2(t1,t2,τG,t0) = ξ(t1,τG,t0)*ξ(t2,τG,t0)
ψ_w = 1/sqrt(2)*twophoton(bw,ξ2,τG,t0)
ψ_in = fockstate(be,0) ⊗ ψ_w

#Calculate output state
sol = waveguide_evolution(times,ψ_in,H)
ψ_out = TwoPhotonView(sol)
\end{minted}
\caption{Same calculation as in Code Sample \ref{ls:single_photon_scattering}, but with an initial two-photon state. Here, the waveguide basis and operators are redefined to include two photons. The two-photon output state can be viewed using the convenience function \code{TwoPhotonView} and is plotted in Fig.~\ref{fig:twophoton_scattering}(a).}
\label{ls:twophoton_scattering}
\end{listing}

\begin{figure}[t]
    \centering
    \includegraphics[width=\linewidth]{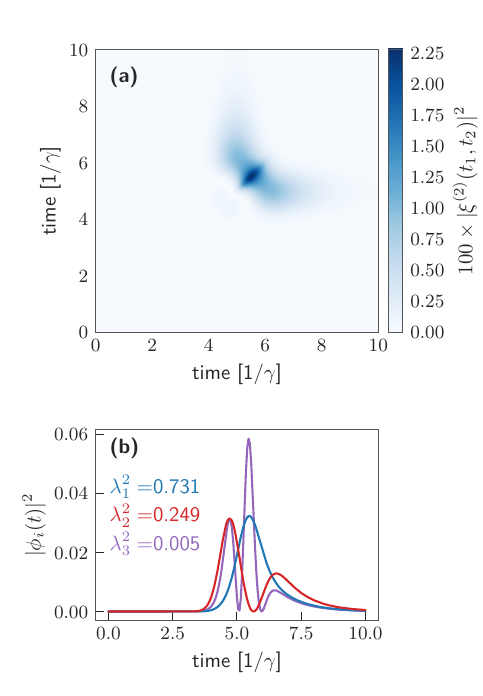}
    \caption{(a) The scattered two-photon wavefunction $\abs{\xi(t,t^\prime)}^2$ produced by Code Sample \ref{ls:twophoton_scattering}. (b) The one-photon decompositions $\phi_i(t)$ of the wavefunction in (a) with the three largest coefficients $\lambda_i^2$.}
    \label{fig:twophoton_scattering}
\end{figure}

We can modify the code in Code Sample \ref{ls:single_photon_scattering} to simulate the scattering of a two-photon Gaussian pulse. That is a two-photon waveguide state as in Eq.~\eqref{eq:two_discretized} with $\xi^{(2)}(t,t^\prime) = \xi^{(1)}(t)\xi^{(1)}(t^\prime)$. This is done in Code Sample \ref{ls:twophoton_scattering} and Fig.~\ref{fig:twophoton_scattering}(a) shows the resulting two-photon output state $\xi_\text{out}^{(2)}(t,t^\prime)$. The Gaussian pulse is reshaped into a complex shape with strong correlations between the two photons. These correlations are evident as long arms or shadows along the diagonal. We further quantify the complexity of the state by expanding the two-photon wavefunction in terms of products of one-photon wavefunctions via a Singular Value Decomposition (SVD) as~\cite{Yang2022DeterministicSystems}:
\begin{equation}
    \xi^{(2)}(t,t') = \sum_i \lambda_i \phi_i(t) \phi_i(t') \label{eq:SVD}
\end{equation}
where $\phi_i(t)$ are orthonormal basis functions, and $\sum_i\lambda_i^2 = 1$ for a normalized wavefunction. For product states (such as the input), $\lambda_1^2 = 1$. In Fig.~\ref{fig:twophoton_scattering}(b), we plot one-photon decompositions $\phi_i(t)$ by including the three largest coefficients $\lambda_i^2$. It is clear that the scattered wavefunction describes an entangled two-photon state comprised of multiple single-photon wavefunctions. Describing two-photon states that are not product states is non-trivial, and in most computational methods, it is either impossible or requires a substantial overhead. In the SLH method, for example, one would have to introduce an additional output channel for each one-photon decomposition one would like to capture \cite{Kiilerich2019Input-OutputPulses,Kiilerich2020QuantumRadiation}. In waveguideQED.jl, the description of entangled two-photon states, on the other hand, comes naturally.

% -------------------------------------------------------
\section{Examples with Multiple Waveguides and non-Markovian Dynamics}
% -------------------------------------------------------
%\todo{is there a more descriptive version of the section title, something like "Examples motivated by ..." or "Examples of nonmarkovian and yadayada dynamics". Maybe use the word "application" or "testcase" instead of "example"}
In this section, we demonstrate and introduce more functionalities of the framework. These include the description of multiple waveguide modes and non-Markovian feedback. More examples can be found in the online documentation~\cite{DocumentationWaveguideQED.jl}.

\subsection{Multiple Waveguides}
\begin{figure*}
    \centering
    \includegraphics[width=\linewidth]{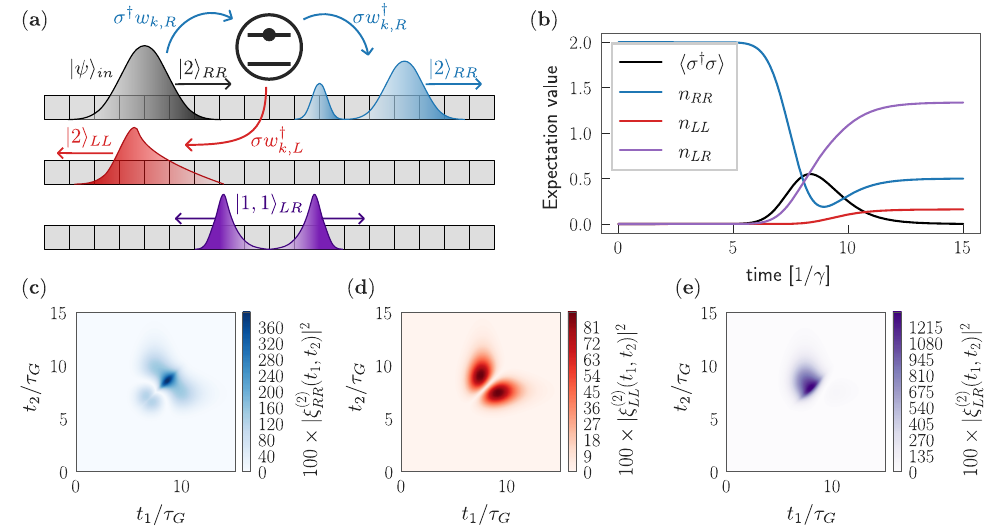}
    \caption{(a): Sketch of the system. A two-photon pulse in the right-propagating mode is scattered into both left- and right-propagating pulses when interacting with a quantum emitter. (b): The populations of the emitter and waveguide modes as a function of time. Here, $n_{XY}(t) = 2 \int \int ds ds^\prime |\xi^{(2)}_{XY}(s,s^\prime)(t)|^2$ and $\xi^{(2)}_{XY}(s,s^\prime)(t)$ is the wavefunction at time $t$. (c)-(e): The three different wavefunctions resulting from the scattering produced by Code Sample \ref{ls:lodahl}.}
    \label{fig:lodahl}
\end{figure*}

\begin{listing}[!ht]
\begin{minted}[
framesep=2mm,
baselinestretch=0.9,
bgcolor=LightGray,
fontsize=\small,
]{julia}
using WaveguideQED,QuantumOptics
#Waveguide basis with two photons and waveguides
times = 0:0.05:15
dt = times[2] - times[1]
bw = WaveguideBasis(2,2,times)

#Waveguide operators of mode 1 and 2
wL,wdL = destroy(bw,1),create(bw,1)
wR,wdR = destroy(bw,2),create(bw,2)

#Define emitter and Hamiltonian
be = FockBasis(1)
s,sd = destroy(be),create(be)
γL, γR = 1/2,1/2 #Total rate will be 1
H = im*sqrt(γL/dt)*(sd ⊗ wL - s ⊗ wdL)+
im*sqrt(γR/dt)*(sd ⊗ wR - s ⊗ wdR)

#Define input state in mode 1 (left)
ξ2(t1,t2,τG,t0) = ξ(t1,τG,t0)*ξ(t2,τG,t0)
τG,t0 = 2,7.5
ψ_w = 1/sqrt(2)*twophoton(bw,1,ξ2,τG,t0)
ψ_in = fockstate(be,0) ⊗ ψ_w
#Calculate output state
sol = waveguide_evolution(times,ψ_in,H)
ψRightScat = TwoPhotonView(sol,1)
ψLeftScat = TwoPhotonView(sol,2)
ψLeftRightScat = TwoPhotonView(sol,1,2)
\end{minted}
\caption{A waveguide basis containing multiple waveguides is defined, and the scattering of a two-photon pulse on a single emitter coupled to two modes is calculated. The resulting wavefunctions are viewed using the convenience function \code{TwoPhotonView} and are plotted in Fig.~\ref{fig:lodahl} (c)-(e).}
\label{ls:lodahl}
\end{listing}

Previously, we considered only a single waveguide mode, but multiple waveguide modes might be desirable to describe, e.g., a right- and left-propagating mode. As a common use case, we consider an emitter placed in the middle of a waveguide. The emitter couples to both the right- and left-propagating modes of the waveguide, and an incident pulse is partly transmitted and partly reflected. Such a system can be realized experimentally using a quantum dot coupled efficiently to a nanophotonic waveguide, as in Ref.~\cite{LeJeannic2022DynamicalEmitter}. Describing multiple waveguides in the framework is trivial, and in the following, we consider the above-mentioned scenario where a two-photon pulse scatters off an emitter coupled to a left- and right-propagating waveguide mode. A sketch of the system is shown in Fig.~\ref{fig:lodahl}(a). The discretized Hamiltonian describing this scenario is:
\begin{equation}
    \begin{aligned}
        H(t) &=  \sum_k f_k(t)\sqrt{\gamma_\mathrm{R}} ( \sigma^\dagger w_{k,\mathrm{R}} + \sigma w^\dagger_{k,\mathrm{R}})  \\ 
        &+ \sum_k f_k(t) \sqrt{\gamma_\mathrm{L}} ( \sigma^\dagger w_{k,\mathrm{L}} + \sigma w^\dagger_{k,\mathrm{L}} ) \label{eq:lodahl_ham}
    \end{aligned}
\end{equation}
where $\gamma_\mathrm{R}$ and $\gamma_\mathrm{L}$ are the decay rates into the right and left propagating waveguide modes, respectively. $w_{k,\mathrm{R}}$ and $w_{k,\mathrm{L}}$ are the annihilation operator of a photon in time-bin $k$ in the right- and left-propagating mode, respectively.

In WaveguideQED.jl, we will, as before, define the discretized waveguide operators, and the time dependence is taken care of behind the scenes. In Code Sample \ref{ls:lodahl}, we show the code necessary to set up a waveguide basis containing two waveguides as well as how to define the Hamiltonian in Eq.~\eqref{eq:lodahl_ham}. We also define an initial two-photon state residing in mode 1 (which we label as the right-propagating mode) and calculate its scattering off the emitter.

%$n_{XY} = 2 \sum_{kj} \abs{\bra{1_{t_k}}_X\bra{1_{t_j}}_Y  \psi(t) \ket{1_{t_k}}_X\ket{1_{t_j}}_Y}^2 $.

In Fig.~\ref{fig:lodahl}(c)-(e), we plot the three possible scattered wavefunctions, $\xi^{(2)}_\mathrm{LL}(t,t^\prime)$, $\xi^{(2)}_\mathrm{RR}(t,t^\prime)$, and $\xi^{(2)}_\mathrm{LR}(t,t^\prime)$, which describe the state $\ket{\psi} = \int \int \xi^{(2)}_{XY}(t,t^\prime) w^\dagger_X(t)w^\dagger_Y(t) \ket{\emptyset}$. The wavefunctions thus represent the case of both photons being in the transmitted right-propagating mode ($\mathrm{RR}$), the reflected left-propagating mode ($\mathrm{LL}$), or one photon in each mode ($\mathrm{LR}$). In Fig.~\ref{fig:lodahl}(b), we also plot the population in the waveguide modes and emitter as a function of time. Note that we can distinguish between the number of photons in the left- ($n_\mathrm{LL}$) or right-propagating mode ($n_\mathrm{RR}$) or the combination, where two photons propagate in opposite directions ($n_\mathrm{LR}$). It is clear that the probability of having both photons in the reflected (left-propagating mode) is very unlikely. This can be understood from the wavefunctions, where the empty anti-diagonal in $\xi^{(2)}_\mathrm{LL}$ also shows that the emitter can only reflect one photon at a time. In $\xi^{(2)}_\mathrm{RR}(t,t^\prime)$, we, on the other hand, see a large probability of having both photons transmitted, especially along the antidiagonal, indicating stimulated emission. In $\xi^{(2)}_\mathrm{LR}(t,t^\prime)$, we see a highly non-symmetric wavefunction indicating a clear time ordering of the events: We will not see a photon in the reflected channel if we did not first observe a photon in the transmitted channel.

\subsection{Non-Markovian Feedback Dynamics}
\begin{figure}[!ht]
    \centering
    \includegraphics[width=\linewidth]{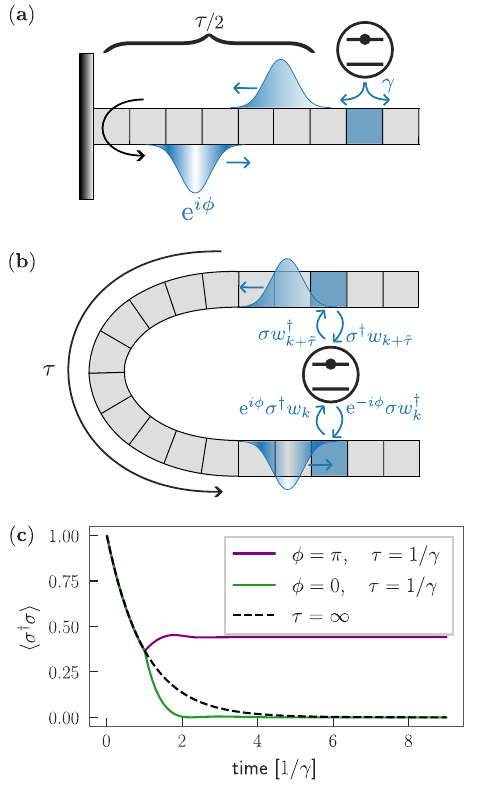}
    \caption{(a): A sketch of a semi-infinite waveguide terminated by a mirror separated from an emitter with a delay $\tau$. The emitted photon is reflected back to the emitter with a phase difference $\mathrm{exp}({i \phi})$. (b): Same as in (a), but the emitter here interacts with the waveguide at two points in time with a phase difference, giving the same effect. This is how the system is set up in WaveguideQED.jl. (c): The emitter population as a function of time for a delay $\tau=1/\gamma$ and two different phases $\phi=0$ and $\phi=\pi$. For reference, the case of $\tau=\infty$ is also shown.}
    \label{fig:delay}
\end{figure}

\begin{listing}[!ht]
\begin{minted}[
framesep=2mm,
baselinestretch=0.9,
bgcolor=LightGray,
fontsize=\small
]{julia}
using WaveguideQED,QuantumOptics
times = 0:0.05:10
dt = times[2] - times[1]
bw = WaveguideBasis(1,times)

#Delay time and operators
τ = 1
w_τ = destroy(bw;delay=τ/dt)
wd_τ = create(bw;delay=τ/dt)
w, wd = destroy(bw), create(bw)

#Define emitter and Hamiltonian
be = FockBasis(1)
s,sd = destroy(be), create(be)
γ,ϕ = 1,pi
H_pi = sqrt(γ/2/dt)*( exp(im*ϕ)*sd ⊗ w 
+ exp(-im*ϕ)*s ⊗ wd + sd ⊗ w_τ + s ⊗ wd_τ)

#Emitter population expectation value
n = sd*s ⊗ identityoperator(bw)
ne_exp(time,psi) = expect(n,psi)

#Initially excited emitter
ψ_in = fockstate(be,1) ⊗ zerophoton(bw)

#Simulate dynamics
ts = 0:0.05:9
_,ne_pi = waveguide_evolution(ts,ψ_in,H_pi;
fout=ne_exp)
\end{minted}
\caption{Delayed waveguide operators are defined and combined with a single-emitter system. The resulting emitter population is plotted in Fig.~\ref{fig:delay}(c).}
\label{ls:delay}
\end{listing}
 
Another powerful example of the framework is its ability to describe systems with memory effects, where the emitted light in the waveguide is fed back into the system. Such memory effects could arise when an emitter is close to a mirror or when two emitters are spatially separated. Waveguide systems with memory effects constitute a challenging class of problems because knowledge of the emitted field is inherently required to capture the correct feedback. As alluded to in the introduction, many numerical approaches such as the SLH \cite{Kiilerich2019Input-OutputPulses,Kiilerich2020QuantumRadiation} or master equation approaches \cite{Baragiola2012N-PhotonSystem} do not describe the emitted field in its entirety. While HEOM \cite{heomfuchs2023,liang2024purifiedinputoutputpseudomodemodel,cirio2025inputoutputhierarchicalequationsmotion} do not include the waveguide field, the correlations are included in the equations allowing for the inclusion of non-Markovian effects. In contrast to the time-binning approach of WaveguideQED.jl, in HEOM, there are fewer restrictions on the spectral density governing the interaction with the waveguide. As also mentioned in the introduction, this comes at the cost of not only increased numerical complexity but also the overall conceptual complexity of the calculation. In WaveguideQED.jl, the restriction of flat-spectral densities allows the very intuitive time-bin picture, which we show in the following, allow for a seamless description of memory effects in waveguide systems.

As an example, we consider a semi-infinite waveguide with a mirror in one end and an emitter placed at a distance from the mirror so that a delay $\tau$ occurs. This is also illustrated in Fig.~\ref{fig:delay}(a), where an emitted photon is reflected by the mirror and accumulates a phase $\phi$ before returning to the emitter. In Fig.~\ref{fig:delay}(b), we sketch an equivalent and more convenient way of viewing this system. Here, the emitter interacts with the waveguide at two points in time that are separated by a delay $\tau$. Due to the moving of the conveyor belt, the photon emitted at the index $k+\Tilde{\tau}$ is seen again by the emitter when $k^\prime=k + \Tilde{\tau}$, where $\Tilde{\tau} = \tau/\Delta t$ is the index necessary to introduce a time-delay of $\tau$. We formulate this problem via the Hamiltonian \cite{Whalen2019CollisionTrajectories}:
\begin{equation}
\begin{aligned}
    H_k &= \sum_k f_k(t) \sqrt{\gamma/2\Delta t} \left( \mathrm{e}^{i \phi} \sigma^\dagger w_{k} + \mathrm{e}^{-i \phi} 
 \sigma w_{k}^\dagger \right) \\
    & + \sum_k f_k(t) \sqrt{\gamma/2\Delta t} \left( \sigma^\dagger w_{k+\Tilde{\tau}} + \sigma w_{k+\Tilde{\tau}}^\dagger \right) \label{eq:tls_feedback}
\end{aligned}
\end{equation}

In Code Sample \ref{ls:delay}, we set up a delayed operator $w_{k+\Tilde{\tau}}$ and the Hamiltonian in Eq.~\eqref{eq:tls_feedback}. We simulate the dynamics of an initially empty waveguide and an excited emitter. In Fig.~\ref{fig:delay}(c), we plot the population of the emitter $\expval{\sigma^\dagger \sigma}$ as a function of time for a phase difference of $\phi=0$ and $\phi=\pi$, respectively. The observed dynamics are very different between the two cases.  For $\phi = 0$, the reflected field interferes constructively with the emitted field into the infinite part of the waveguide.  This leads to a faster decay of the emitter, which is evident by comparing with the reference case of $\tau=\infty$, which is also shown. With $\phi = \pi$, we instead see destructive interference, and the excitation can never escape into the infinite part of the waveguide. This leads to the excitation being trapped (a cavity is formed between the mirror and emitter), and the system ends up in a steady state where the emitter has a constant population of $\expval{\sigma^\dagger \sigma}\approx 0.5$.

% -------------------------------------------------------
\section{Performance, Limitations, and Outlook \label{sec:performance}}
% -------------------------------------------------------
In this section, we discuss the performance of WaveguideQED.jl. Specifically, we emphasize how the implementation exploits the special structure of the time-dependent waveguide operators to have matrix-free multiplication and also represents the waveguide basis in the most efficient way. We show that this implementation is orders of magnitude more efficient than naively allocating a sparse matrix for each time-bin operation and also show a significant speed up when compared with a similar space-discretized waveguide method \cite{ArranzRegidor2021ModelingModel}. We also discuss the current limitations of the framework, their impact, and the possibility of extending its capabilities. 

\subsection{LazyOperators}
In Fig.~\ref{fig:conveyor}(b), we illustrate how the time-binned waveguide operator $w_k^\dagger$ acts on the waveguide states. The structure is quite clear; we are only addressing time-bins with indices $k$. We exploit this structure in the implementation of waveguide operators and have the transformation that occurs due to $w_k^\dagger$ be performed not by a matrix but by an equivalent-but-faster computation kernel function, where we can change the time index $k$ effortlessly and efficiently. This allows us to change the time-bin index that the waveguide operators act on during the simulation by simply changing the input $k$ to the kernel function and thus performing the evolution sketched in Fig.~\ref{fig:conveyor}(c).  This, however, complicates things when operators need to be combined either via products, sums, or tensor products to form a desired Hamiltonian since there is no actual matrix on which to perform these operations.  
%This is trivial when we only consider a waveguide state, but when combined with other Hilbert spaces, we can no longer simply perform the tensor product to get the waveguide operator acting on the new combined Hilbert space.

Instead, we need to use LazyOperators \cite{LazyOperatorsQuantumOptics.jl}, which we explain with some basic examples in the following. The general idea of LazyOperators is to delay the evaluation of the sum, product, or tensor product until the operator is actually applied on a state. As such, every LazyOperator is an object with a list of operators that it needs to apply when evaluated. For example, the LazySum, implementing a delayed sum, can be understood from the following simplified Julia code:
\begin{listing}[H]
\begin{minted}[
framesep=2mm,
baselinestretch=0.9,
bgcolor=LightGray,
fontsize=\small
]{julia}
C = LazySum(A,B) # C = A + B
#Function to compute ψ_out = C*ψ_in
function mul!(ψ_out,C::LazySum,ψ_in)
    for operator in C.operators
        ψ_out += operator * ψ_in
    end
    return ψ_out
end
\end{minted}
\end{listing}
Here, we create an object \code{C}, which contains the operators \code{A} and \code{B} saved in \code{C.operators}. When \code{C} is applied to a state \code{ψ\_in}, we loop over all operators in  \code{C.operators} and apply them to \code{ψ\_in} and while adding the result to \code{ψ\_out}. At the end of the loop, we have then computed $\psi_{out} = \sum_i C_i \psi_{in}$, where $C_i$ is here \code{C.operators[i]}.

Similarly, a lazy product would be implemented as:
\begin{listing}[H]
\begin{minted}[
framesep=2mm,
baselinestretch=0.9,
bgcolor=LightGray,
fontsize=\small
]{julia}
C = LazyProduct(A,B) # C = A*B
#Function to compute ψ_out = C*ψ_in
function mul!(ψ_out,C::LazyProduct,ψ_in)
    ψ_out = operator * ψ_in
    for operator in C.operators[end-1:-1:1]
        ψ_out = operator * ψ_out
    end
    return ψ_out
end
\end{minted}
\end{listing}
Here, again \code{C} contains the operators \code{A} and \code{B} in \code{C.operators}, and when applied we loop (in reverse) through all the operators and apply them to $\psi_{in}$ to compute in the end: $\psi_{out} = \left ( \prod_{i=N}^{i=1} C_i \right) \psi_{in}$, where again $C_i$ is \code{C.operators[i]} and $N$ is the number of operators and the product runs in reverse. 

The LazyTensor is by far the most complicated operation, and to explain how it works better, we start with a simple example. We consider the tensor product between two two-level emitters:
\begin{equation}
     \begin{pmatrix} {a} \\ {b} \end{pmatrix} \otimes \begin{pmatrix} {c} \\ {d} \end{pmatrix} = \begin{pmatrix} {a}{c} \\ {a} {d} \\ {b}{c} \\ {b}{d} \end{pmatrix}
\end{equation}
If, we now want to apply $\sigma = \begin{pmatrix}
    0 & 1 \\ 0 & 0 \end{pmatrix}$ to the second two-level system subsystem, we could do that by:
\begin{equation}
    I \otimes \sigma \begin{pmatrix} {a}{c} \\ {a}{d} \\ {b}{c} \\ {b}{d} \end{pmatrix} = \begin{pmatrix}
    \color{red}{\begin{pmatrix}
    0 & 1 \\ 0 & 0
\end{pmatrix}} \\ \color{blue}{\begin{pmatrix}
    0 & 1 \\ 0 & 0
\end{pmatrix}}
\end{pmatrix} \begin{pmatrix} \color{red}{a}{c} \\ \color{red}{a}{d} \\ \color{blue}{b}{c} \\ \color{blue}{b}{d} \end{pmatrix} = \begin{pmatrix} \color{red}{a}{d} \\ 0 \\ \color{blue}{bd} \\ 0 \end{pmatrix}
\end{equation}
It is here understood that the red matrix is applied to red elements and the blue matrix to blue elements. It is easy to check that this is indeed the same as calculating the matrix $I \otimes \sigma$ and applying the resulting matrix. The above example shows that a tensor product can also be achieved by applying the operators to appropriate subspaces of the total Hilbert space. This is the principle behind LazyTensor. Formally, the code could look something like:
\begin{listing}[H]
\begin{minted}[
framesep=2mm,
baselinestretch=0.9,
bgcolor=LightGray,
fontsize=\small
]{julia}
C = LazyTensor(A,B) # C = A ⊗ B
#Function to compute ψ_out = C*ψ_in
function mul!(ψ_out,C::LazyTensor,ψ_in)
    mul_sub!(ψ_out,C.operators[1],C.loc[1],ψ_in)
    for i in 2:length(C.operators)
        mul_sub!(ψ_out,C.operators[i],
                 C.loc[i],ψ_out)
    end
    return ψ_out
end
\end{minted}
\end{listing}
Here, the function is very similar to the Lazyproduct function, except we use the function \code{mul\_sub!} to apply the operator appropriately on the subspaces based on the which subspace the operator belongs to, which is stored in C.loc. We do not include the code for \code{mul\_sub!} as it is beyond the scope of this quick introduction to LazyOperators.  

With the above three types of LazyOperators and with some rules for how they combine, we can have the efficiency of the matrix-free kernel function while at the same time being able to combine operators to create arbitrarily complex and general Hamiltonians.
%instead of performing the tensor product, save all the operators involved and sequentially apply them to the appropriate sub-Hilbert spaces. The Lazy Operators are thus semi-symbolic objects, where the tensor product (or other algebraic operations) are represented in a structured way that can be exploited for fast application.

\subsection{Benchmarks}
\begin{figure}
    \centering
    \includegraphics[width=\linewidth]{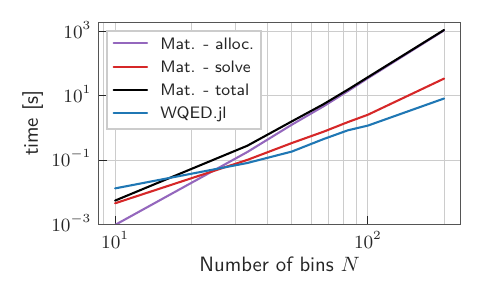}
    \caption{The computational time of processes involved in solving the problem in Code Sample \ref{ls:twophoton_scattering} as a function of the number of bins. Larger amount of bins means smaller dt. The last data point corresponds to $dt=0.05$ as in Code Sample \ref{ls:twophoton_scattering}. Mat. is the naive matrix method that allocates sparse matrices before solving the problem. Notably, WaveguideQED.jl is orders of magnitude faster than the matrix method. }
    \label{fig:benchmark}
\end{figure}

To benchmark our implementation, we compare it with the naive implementation of allocating a sparse matrix for the Hamiltonian $H_k$ for each time index $k$ in the simulation. The matrix for the waveguide operator of time-bin 3 $w_3^\dagger$ in Fig.~\ref{fig:conveyor}(b) would be given as:
\begin{equation}
    w_3^\dagger \ket{\psi} = \begin{pmatrix}
        0 & 0 & 0 & 0 & 0 & 0 \\
        0 & 0 & 0 & 0 & 0 & 0 \\
        0 & 0 & 0 & 0 & 0 & 0 \\
        1 & 0 & 0 & 0 & 0 & 0 \\
        0 & 0 & 0 & 0 & 0 & 0 \\
        0 & 0 & 0 & 0 & 0 & 0 \\
    \end{pmatrix} \cdot \begin{pmatrix}
        \textcolor{red}{\psi_\emptyset} \\
        \psi_1 \\
        \psi_2 \\
        \psi_3 \\
        \psi_4 \\
        \psi_5
    \end{pmatrix} = \begin{pmatrix}
        0 \\
        0 \\
        0 \\
        \textcolor{red}{\psi_\emptyset} \\
        0 \\
        0
    \label{eq:operator_matrix} \end{pmatrix} 
\end{equation}
here, $\psi_i$ denotes the value stored in element $i$ (starting from zero to represent vacuum and the subsequent $N$ elements to represent the single photon excitation). In the benchmark, we then preallocate $N$ sparse matrices corresponding to the waveguide operators $w_1...w_N$ and then have a time-dependent Hamiltonian that switches between each waveguide operator as time progresses. As $N$ increases, the memory necessary to store all these matrices increases substantially, and even worse, the time it takes to preallocate said matrices can become unacceptable. In Fig.~\ref{fig:benchmark}, we show a benchmark of the computational time involved in first allocating the matrices and then solving the differential equation. We vary the number of bins in the waveguide state (similar to changing $\Delta t$), and a larger amount of bins means a smaller $\Delta t$. We perform the benchmark on the two-photon pulse scattering example considered in Code Sample \ref{ls:twophoton_scattering} and displayed in Fig.~\ref{fig:twophoton_scattering}. We also show the computational time of WaveguideQED.jl, which involves both creating the Lazy Operators (very cheap) and solving the differential equation. From the figure, it is clear that as the number of bins increases and thus the accuracy of the time-binning of photons, the computational cost of just allocating matrices far outweighs the time it takes to solve the differential equation. Notably, WaveguideQED.jl is orders of magnitude faster than the total compute time of the matrix method while using significantly less memory since the memory usage of WaveguideQED.jl is constant in the number of time-bins $N$ due to the matrix-free kernel function underneath. The sparse matrices, on the other hand, scale as $O(N^3)$, and for $dt=0.1$, they use four orders of magnitude more memory than WaveguideQED.jl. The example in Code Sample \ref{ls:twophoton_scattering}, which uses $\Delta t=0.05$ corresponding to the last data point, would thus have taken over half an hour using matrices but takes less than 10 seconds using WaveguideQED.jl. Moreover, even if we do not count the significant memory-allocation time overhead, the structured computational kernel function used for operator application in WaveguideQED.jl is much faster than the multiplication by a sparse matrix.

To further benchmark our implementation, we compare with the simulation times in ref.~\cite{ArranzRegidor2021ModelingModel}. Here, they employ a similar time-bin picture as in WaveguideQED.jl to simulate the feedback system considered in Fig.~\ref{fig:delay}. They, however, only describe the part of the waveguide between the emitter and mirror. Photons emitted to the right of the emitter (into the infinite part of the waveguide) are not included, but instead, a "measurement" is performed on them, similar to having a quantum jump. The evolution is then calculated by realizing many such Monte-Carlo trajectories. Depending on the number of trajectories considered, the simulation time of the system considered in Fig.~\ref{fig:delay} is around 1-30s on a computer workstation. For comparison, the same simulation (with the same parameters) takes around 70ms using WaveguideQED.jl on a laptop with an i5-1345U, 1.6GHz, 10-core, processor. While this might not be a fundamental algorithmic advantage, it does demonstrate the efficiency and high performance of the WaveguideQED.jl implementation.      

\subsection{Efficient Choice of Basis}
In this section, we briefly discuss how we use a custom basis in WaveguideQED.jl to represent the waveguide states efficiently, which is conceptually similar to Excitation Number Restricted States in QuTiP5 \cite{Lambert2024}. This smart basis is also a large factor in what allows the code to run fast. To appreciate the smart basis choice, let us start by considering a naive basis choice. If the one photon waveguide state consists of $N$ bins, the local system interacts with $N$ different bosonic modes. Limiting each mode to a maximum of one photon would make the total number of elements in the waveguide state $2^N \approx 1.26 \cdot 10^{30}$. This is simply infeasible, but we are also here allowing up to $N$ excitations to be present simultaneously in the waveguide. Instead, by restricting the system to a single excitation, we only need to include the vacuum state \(\ket{\emptyset} = \ket{0}_1 \otimes \ket{0}_2 \otimes \cdots \otimes \ket{0}_N\) and single excitations in mode $k$: \(\ket{1_k} = \ket{0}_1 \otimes \cdots \otimes \ket{1}_k \otimes \cdots \otimes \ket{0}_N\). Here, \(\ket{0}_i\) and \(\ket{1}_k\) represent the vacuum or a single excitation in the \(i\)-th mode. This way, the total number of elements in the basis is reduced to \(N+1\), corresponding to the vacuum state and the states with a single excitation in each mode. 

In a similar manner, we efficiently represent two-photon states and multiple waveguide states by limiting the total number of excitations to two. For a two-waveguide state, we thus use the basis states $\ket{1_k 1_j} = \ket{0}_1 \otimes \cdots \otimes \ket{1}_k \otimes \cdots \otimes \ket{1}_j \otimes \cdots \otimes \ket{0}_N$. Since $\ket{1_k 1_j} =\ket{1_j 1_k}$, this gives a total of $N(N+1)/2$ elements. When we have multiple waveguide $N_W$, then we introduce basis states $\ket{1_k}_A \ket{1_j}_B$, where $\ket{1_k}_A$ denotes an excitation in time-bin $k$ of waveguide A. Here, $\ket{1_k}_A \ket{1_j}_B \neq \ket{1_j}_A \ket{1_k}_B$ and thus we need $N^2$ elements per waveguide pairing. The number of pairings for $N_W$ waveguides is $N_W(N_W-1)/2$, which gives a total of $N_W N(N-1)/2 + N_W N^2$ number of elements, which again is much more favorable than $(N(N-1)/2)^{N_W}$.

%\todo{I think a section on the fact that you have single and two-photon special bases is pretty important and deserves its own section here -- it is a non-obvious way to make the code faster (or even possible)}

\subsection{Limitations}

The most obvious limitation of the framework is the stringent requirement of a maximum of two photons. This limitation is not fundamental, and it is easily possible to extend the framework to more photons. Implementation-wise, such an extension would be rather straightforward, but its computational cost might be too high for any practical simulation. Due to the symmetry of a bosonic wavefunction, we need $\frac{(N+n-1)!}{n!(N-1)!}$ elements to describe an $n$-photon state with a time-discretization of $N$ bins. For $N=100$ bins with $n=2$, this amounts to $\approx 5 \cdot 10^4$ elements, with $n=3$ its $\approx 1.7 \cdot 10^5$ elements, and for $n=4$ its $4.4 \cdot 10^6$ elements. For comparison, the simulation time for $n=2$ with $N=100$ bins is $\approx 1 \ \mathrm{s}$, and assuming linear scaling in computational time per element number, this would mean $\approx 34 \ \mathrm{s}$ and $\approx 875 \ \mathrm{s}$ for $n=3$ and $n=4$, respectively. These are still reasonable simulation times, but anything beyond this is unrealistic, and some of the more restrictive numerical schemes such as the SLH \cite{Kiilerich2019Input-OutputPulses,Kiilerich2020QuantumRadiation,Lund2023PerfectPulse,Yang2022DeterministicSystems} or more complex schemes such as matrix product states \cite{Pichler2016PhotonicFeedback,Pichler2017UniversalFeedback,Crowder2020QuantumFeedback,Crowder2022QuantumFeedback,ArranzRegidor2021ModelingModel,Richter2022EnhancedNetworks} are probably more suitable for such studies. We note that the possibility of using a Graphical Processing Unit (GPU) for particularly demanding tasks is readily available using CUDA.jl and custom-written GPU kernel functions available in the WaveguideQED.jl framework (for more details, see the documentation \cite{DocumentationWaveguideQED.jl}). This could also help extend the domain of numerically feasible tasks.  

At this point, it is also worth questioning whether larger photon states are what one would study using the formalism and framework presented here. Many of the fundamental photon-photon interactions can be studied with just a two-photon description. We have focused on the scattering of single emitters, but there are, in principle, no limitations to what local quantum systems can be studied in our framework. Another prime example of photon-photon interactions is the beamsplitter, which is also readily available with a two-photon description and, in fact, has also been studied using the framework; see the online documentation \cite{DocumentationWaveguideQED.jl} for more details. The main advantages of the framework are thus its high performance, flexibility, generality, ease of use, and ability to be combined with any complex local quantum systems.%\todo{Any idea whether autodiff works out of the box? The potential support for it should definitely be mentioned}

\subsection{Outlook}
While we have already shown that WaveguideQED.jl can tackle a variety of different problems, it is still relatively early in its development, and there are plenty of directions the framework can take. In this section, we discuss possible future additions and studies that are worth pursuing with the framework.

In the examples presented here, no losses to the environment have been considered. Adding Lindblad loss terms to the simulations is trivial, albeit leading to larger numerical overhead. Since the waveguide states are implemented as state vectors and not density matrices, such an addition would most naturally utilize the Monte Carlo wavefunction approach. Here, multiple trajectories would be computed, thus raising modestly the computational costs. In return, one would be able to study one and two-photon wavefunctions subject to losses, which is a relatively unexplored regime due to the complexity of the calculations involved. Even in many matrix-product states, such losses are not easily included \cite{ArranzRegidor2021ModelingModel}. Implementation-wise, there are no notable additions for such an extension of the framework, and such studies would be readily available; it is just a matter of changing the numerical ODE solver.

The framework could also be extended to larger photon states. This would enable studies of three and four-photon wavefunctions, which could provide valuable insights into the dynamics of higher-order photonic states. This direction is, however, limited due to the numerical considerations also discussed in the last section.

Finally, more exotic waveguide systems, such as systems with non-linear dispersion \cite{PhysRevA.102.023702}  or multiple contact points/giant emitters \cite{PhysRevResearch.2.043070} could be studied using the framework.

% -------------------------------------------------------
\section{Conclusion}
% -------------------------------------------------------
We have developed a fast, intuitive, easy-to-use, and general simulation tool for waveguide quantum electrodynamics. In the framework, we represent traveling multi-photon states as a collection of time-binned modes, represented in a well-chosen custom basis, interacting with a local system. This allows for a clear mental picture of the propagation of the photonic state. The waveguide states and operators can easily be combined with states and operators defined using QuantumOptics.jl, allowing the simulation of arbitrarily complex local systems interacting with waveguides. 

We demonstrated the efficiency of the software, which uses a matrix-free implementation of the time-dependent waveguide operators. This implementation exploits the structure of waveguide operators to significantly reduce the numerical cost of applying them. This is possible due to the implementation of Lazy Operators in QuantumOptics.jl, which we show leads to significant performance gains. 

The package is an open-source project still in development, and we encourage community feedback, bug reports, and feature requests.

\begin{acknowledgments}
This work was supported by the Danish National Research Foundation (NanoPhoton - Center for Nanophotonics, Grant No. DNRF147) and Villum Fonden (QNET-NODES grant no. 37417).
\end{acknowledgments}

\onecolumn\newpage
\appendix

% -------------------------------------------------------
\section{Convergence \label{app:single_photon}}
% -------------------------------------------------------
In sec. \ref{sec:theory}, we derived the equations of motion for a single photon pulse scattering on a single emitter. In this Appendix, we solve the equations of motion analytically and check the convergence of WaveguideQED.jl.

For a Gaussian pulse as defined in Code Sample \ref{ls:single_photon_scattering}, we can solve the differential equation in eqs.~\eqref{eq:eom1} and \eqref{eq:eom2} as:
\begin{align}
    &a = \frac{2 \log(2)}{\tau_G^2}, \ \ b = \frac{\gamma}{2}, \ \ x = 2a(t - t_0) - b, \ \  y = 2a t_0 + b \\
    &I(t) = \sqrt{\pi} \frac{\exp\left(\frac{b^2}{4a} + bt_0\right)}{2\sqrt{a}} \left( \text{erf}\left(\frac{x}{2\sqrt{a}}\right) + \text{erf}\left(\frac{y}{2\sqrt{a}}\right) \right) \\
    &\psi_e(t) = \sqrt{\frac{2\gamma}{\tau_G}} \left(\frac{\log(2)}{\pi}\right)^{\frac{1}{4}} \exp\left(-\frac{\gamma}{2} t \right) I(t) \\
    &\xi_\text{out}(t) = \xi_{in}(t) - \sqrt{\gamma}\psi_e(t) \label{eq:eom_sol}
\end{align}
with $\mathrm{erf}$ being the error function. 
We now compare the solution obtained from Code Sample \ref{ls:single_photon_scattering} with the solution in Eq.~\eqref{eq:eom_sol} and vary the number of bins used to discretize the waveguide (or equivalently reducing the time step $dt$). To quantify the error, we use the L-2 Norm:

\begin{equation}
    ||x||_2 = \int |x(t)|^2 dt 
    \label{eq:l2norm}
\end{equation}

\begin{figure}
    \centering
    \includegraphics[width=0.5\linewidth]{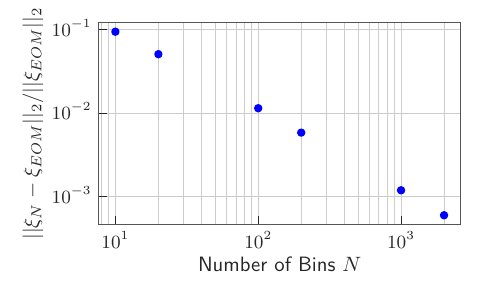}
    \caption{The relative error of the L2-norm between the EOM solution and WaveguideQED.jl solution as defined in Eq.~\eqref{eq:l2norm}.}
    \label{fig:convergence}
\end{figure}

In Fig.~\ref{fig:convergence}, we plot the relative error of the L-2 norm, where $\xi_N$ is the solution of WaveguideQED.jl for $N$ time-bins and $\xi_{EOM}$ is the EOM solution in eq.\eqref{eq:eom_sol}. We see that as the number of bins increases, the relative error decreases accordingly. 

\end{document}